\newif\ifCorr
\Corrtrue          %

\ifCorr
 \documentclass[11pt]{article}
 \usepackage{chicagor}
\else
 \documentclass[10pt,twocolumn]{article}
 \usepackage{latex8}
 \newcommand{\citeyear}[1]{\cite{#1}}
\fi

 \usepackage{times}
 \usepackage{url}
 \usepackage{amsfonts}
 \usepackage{latexsym} %

 \usepackage{amsmath}

\ifCorr

\fi

\newcommand{\thmcolon}{\hspace{-.55em}{\bf :}}
\newtheorem{THEOREM}{Theorem}[section]
\newenvironment{theorem}{\begin{THEOREM} \thmcolon  }%
                        {\end{THEOREM}}
\newtheorem{LEMMA}[THEOREM]{Lemma}
\newenvironment{lemma}{\begin{LEMMA} \thmcolon  }%
                      {\end{LEMMA}}
\newtheorem{COROLLARY}[THEOREM]{Corollary}
\newenvironment{corollary}{\begin{COROLLARY} \thmcolon  }%
                          {\end{COROLLARY}}
\newtheorem{PROPOSITION}[THEOREM]{Proposition}
\newenvironment{proposition}{\begin{PROPOSITION} \thmcolon  }%
                            {\end{PROPOSITION}}
\newtheorem{DEFINITION}[THEOREM]{Definition}
\newenvironment{definition}{\begin{DEFINITION} \thmcolon  \rm}%
                            {\end{DEFINITION}}
\newtheorem{CLAIM}[THEOREM]{Claim}
                            {\end{CLAIM}}
\newtheorem{EXAMPLE}[THEOREM]{Example}
\newenvironment{example}{\begin{EXAMPLE} \thmcolon  \rm}%
                            {\end{EXAMPLE}}
\newtheorem{REMARK}[THEOREM]{Remark}
\newenvironment{remark}{\begin{REMARK} \thmcolon  \rm}%
                            {\end{REMARK}}

\newcommand{\thm}{\begin{theorem}}
\newcommand{\lem}{\begin{lemma}}
\newcommand{\pro}{\begin{proposition}}
\newcommand{\dfn}{\begin{definition}}
\newcommand{\rem}{\begin{remark}}
\newcommand{\xam}{\begin{example}}
\newcommand{\cor}{\begin{corollary}}
\newcommand{\prf}{\noindent{\bf Proof:} }
\newcommand{\ethm}{\end{theorem}}
\newcommand{\elem}{\end{lemma}}
\newcommand{\epro}{\end{proposition}}
\newcommand{\edfn}{\bbox\end{definition}}
\newcommand{\erem}{\bbox\end{remark}}
\newcommand{\exam}{\bbox\end{example}}
\newcommand{\ecor}{\end{corollary}}
\newcommand{\eprf}{\bbox\vspace{0.1in}}
\newcommand{\beqn}{\begin{equation}}
\newcommand{\eeqn}{\end{equation}}
\newcommand{\bbox}{\vrule height7pt width4pt depth1pt}

\newenvironment{oldthm}[1]{\medskip\par\noindent{\bf Theorem #1:} \em \noindent}{\par}
\newenvironment{oldlem}[1]{\medskip\par\noindent{\bf Lemma #1:} \em \noindent}{\par}
\newenvironment{oldcor}[1]{\medskip\par\noindent{\bf Corollary #1:} \em \noindent}{\par}
\newenvironment{oldpro}[1]{\medskip\par\noindent{\bf Proposition #1:} \em \noindent}{\par}
\newcommand{\othm}[1]{\begin{oldthm}{\ref{#1}}}
\newcommand{\eothm}{\end{oldthm} \medskip}
\newcommand{\olem}[1]{\begin{oldlem}{\ref{#1}}}
\newcommand{\eolem}{\end{oldlem} \medskip}
\newcommand{\ocor}[1]{\begin{oldcor}{\ref{#1}}}
\newcommand{\eocor}{\end{oldcor} \medskip}
\newcommand{\opro}[1]{\begin{oldpro}{\ref{#1}}}
\newcommand{\eopro}{\end{oldpro} \medskip}
\newcommand{\sat}{\models}
\newcommand{\rimp}{\Rightarrow}
\newcommand{\riff}{\Leftrightarrow}
\newcommand{\nthm}[1]{\begin{oldthm}{#1}}
\newcommand{\enthm}{\end{oldthm} \medskip}

\newcommand{\sep}{~:~}

\newcommand{\inter}{\cap}
\renewcommand{\phi}{\varphi}
\newcommand{\intension}[1]{[\![ #1 ]\!]}

\newcommand{\commentout}[1]{}

\newcommand{\cL}{\mathcal{L}}

\newcommand{\cP}{\mathcal{P}}

\newcommand{\cU}{\,\mathcal{U}}

\newcommand{\R}{\mathbb{R}}
\newcommand{\N}{\mathbb{N}}

\newcommand{\llic}{\mathcal{L}^{\mathit{lic}}}

\ifCorr
  \title{A Logic for Reasoning about Digital Rights\thanks{This paper
is essentially the same as one that appeared in the \emph{Proceedings
of the 15th IEEE Computer Security Foundations Workshop},
pp. 282--294, 2002.}}
  \author{Riccardo Pucella\\
  Cornell University\\Ithaca, NY 14853\\
  riccardo@cs.cornell.edu
  \and
  Vicky Weissman\\
  Cornell University\\Ithaca, NY 14853\\
  vickyw@cs.cornell.edu}
  \date{}
\else

\title{A Logic for Reasoning about Digital Rights}
\author{Riccardo Pucella\\[.1in]Department of Computer Science\\Cornell
University\\Ithaca, NY 14853\\
riccardo@cs.cornell.edu
\and
Vicky Weissman\\[.1in]Department of Computer
Science\\Cornell University\\Ithaca, NY 14853\\
vickyw@cs.cornell.edu}

\fi

\begin{document}
\maketitle

\ifCorr
\else
 \thispagestyle{empty}
\fi

\begin{abstract}  
We present a logic for reasoning about licenses, which are ``terms of use'' for 
digital resources.  The logic provides a language for writing both properties of 
licenses and specifications that govern a client's actions.  We discuss the complexity 
of checking properties and specifications written in our logic and propose a technique 
for verification.  A key feature of our approach is that it is essentially 
parameterized by the language in which the licenses are written, provided that this language 
can be given a trace-based semantics. We consider two license languages to illustrate this 
flexibility. 
\commentout{
A key feature of our approach is that it is essentially parameterized by the language 
in which the licenses are written, provided that this language can be given a trace-based 
semantics. We consider a few license languages to illustrate this flexibility.  We then 
address the problem of checking properties and specifications written in our logic, by 
proposing techniques for verification.
}
\end{abstract}

\section{Introduction}
In the world of digital rights management, licenses are agreements
between the distributors and consumers of digital resources.  A license
is issued by an owner to a prospective client.  It states the exact 
conditions under which a particular resource may be used, including a 
complete description of how compensation may be given.  Licenses can be 
viewed as a subset of authorization policies, policies that dictate what 
actions a system's principal can perform at any given time.  Licenses 
are an essential part of any rights management system, because they tell 
the consumer,  as well as the enforcement mechanism, which uses are 
legitimate. 
Licenses must be written in some language.  Although
many licenses are very simple 
(e.g., ``consumer must pay a fee before each access to an on-line journal''), 
more complicated ones, in particular ones
involving time, are also common 
(e.g., ``for each month from 1/1/01 to 1/1/02 the mortgage requires either 
a \$1500 payment between the first and fourth of the month or a \$1525 
payment between the fourth and the fourteenth'').  
The language must be expressive enough to capture these
types of licenses.  Languages such as \emph{DPRL} \cite{Ramanujapuram98},
\emph{XrML} \cite{XrML00}, and \emph{ODRL} \cite{ODRL01} have been
developed to state a wide range of licenses. These languages, however,
do not have formal semantics.  Instead, they rely on intuitions behind
their syntax, and on informal descriptions of expected behavior.  As a
consequence, licenses that ``seem right'' are enforced without anyone
knowing precisely what is intended or exactly what is allowed.

\commentout{
To remove these ambiguities, Gunter \emph{et al.~} \citeyear{Gunter01}
used techniques from programming language semantics \cite{Hoare85} to
develop a simple language and a corresponding semantics to express
licenses.  Their approach can be summarized as follows.  The meaning of
a license is a set of traces. Each trace represents a sequence of actions
allowed by the license.  A correct enforcement mechanism permits any 
sequence of action specified by the license and forbids any other.  Using 
the language of Gunter \emph{et al.}\citeyear{Gunter01}, we can state a 
number of licenses precisely.
}
Gunter \emph{et al.} \citeyear{Gunter01}
used techniques from programming language semantics 
\cite{Hoare85} to remove these ambiguities.  In their approach, the meaning 
of a license is a set of traces. Each trace represents a sequence of 
actions allowed by the license.  A correct enforcement mechanism permits 
any sequence of action specified by the license and forbids any other.  To 
illustrate
their idea, Gunter \emph{et al.} 
defined
a simple language with semantics that could be used to state a number
of licenses precisely. 

In addition to unambiguously expressing licenses, we would like to reason about 
them.  
\commentout{
In other words, we want to verify that a license has certain properties. Examples 
of properties include ``is a type-B user ever allowed to view the high-resolution 
image'' and ``is a user ever allowed to access a work without, at some time, 
being obligated to pay.''  Depending on the license, each property may or may not 
be easy to check.  This verification is important, because it increases our 
confidence that the license matches the informal requirements and that the 
informal requirements match the owner's intent.  
}
In general, we are interested in two classes of questions: 
does a set of licenses have certain properties and does a client's actions with respect to 
a set of licenses meet particular specifications. 
Note that we make a distinction 
between the characteristics inherent in a set of licenses 
(properties, sometimes referred to as license properties for emphasis).
and those whose truth depends on the client's actions 
(specifications, sometimes referred to as client behavior specifications for emphasis)
Examples of properties
include 
``a religious work may only be viewed during the hour before sunset''
and ``if a user
accesses a work, then the user is obligated to pay for the access at some time.''  Depending on
the licenses, each property may or may not be easy to check.  Continuing the last example, an owner 
may allow a client to defer payment in so many situations that it is not clear that there will
ever be an occasion when the client must pay.  Alternatively, a license may permit free access to 
some resources, however, the license has so much ``red tape'' that the client cannot determine if 
the desired 
resource
actually
is free.  As for specifications, examples include ``the client never uses a 
resource illegally'' and ``the client is never obligated to pay interest on her credit card debt''.  
The difficulty of specification checking is based on the licenses and the client's actions.  Verifying 
properties and specifications is important, because it increases our confidence that the licenses 
match the informal requirements and that the informal requirements match the owner's intent.  

\commentout{
In this paper we present a logic for reasoning about licenses.  Why a logic? A 
logic provides us with a language in which we can write formal \emph{specifications} 
and  prove rigorously that these specifications are met by a set of licenses.  
\footnote{As such, the relationship between Gunter \emph{et al.}'s work and ours is 
analogous to the one between programming language semantics and Dynamic Logic
\cite{Harel00}.} In addition, automated tools can be developed to verify, through a 
process called \emph{model checking} \cite{Clarke99}, that an application does not 
violate the specifications.  Note that the language in which we write our 
specifications is distinct from the language in which we write the licenses.}

\commentout{
In this paper we present a logic for reasoning about licenses, that
provides us with a language, distinct from the one used to write
licenses, in which 
we
can state properties and specifications
precisely. The logic, essentially a temporal logic, allows us to make
statements about licenses issued, 
assuming the licenses are written in some particular language.  (For ease of 
exposition, we 
use a language of regular expressions to state licenses.)
}
In this paper we present a logic for reasoning about licenses that provides us with
a language in which we can state properties and specifications precisely.  The logic
is essentially a temporal logic.  It allows us to make statements
about issued licenses,
assuming the licenses are written in some particular language
that is distinct from our logic.
For ease of exposition, we assume 
until Section~\ref{s:gunter} that 
licenses are written in a very simple, regular language 
and that the application has only one client and one provider.  Our framework can be 
modified in a straightforward manner to reason about different license languages.  It is 
also easy to extend the logic to multiple clients and providers.  

\commentout{
As we notice by looking at the previous examples,
properties of licenses and specifications typically involve permissions and
obligations of the client to perform certain actions. We take a very
simple view of permissions and obligations in this
paper. 
In particular, we focus exclusively on the client's view of 
the interaction between the provider and the client. 
Inspired by Gunter \emph{et al.}, we interpret licenses as specifying a 
set of legal sequences of actions. A client is permitted to perform an 
action at a given point in time if that action is part of a sequence of 
actions that is legal according to the licenses issued. Similarly, a client 
is obligated to perform an action at a given point in time if that action 
is the only action that can be performed legally by the client
according to the actions it has already performed and the licenses
issued.
}
As the examples suggest, license properties and client behavior specifications
typically involve the client's permissions and obligations to do certain actions.
We take a very simple view of permissions and obligations.
In particular, we focus exclusively on the client's viewpoint.
Inspired by Gunter \emph{et al.}, we interpret licenses as 
describing a set of legal sequences of actions.  A client is permitted to do an action 
if that action is part of a sequence of actions that is legal according to the actions 
she has already done and the licenses issued.  If there is only one such action for a 
particular license, then the client is obligated to do that action.  

\commentout{
As an example, consider a license for a mortgage. The mortgage agreement
requires the client to sign the appropriate documents and then to make a 
\$3000 payment every six months or a \$6000 payment every year.  For the 
first year, there are two legal action sequences.  The client could sign 
the lease, pay \$3000 after the first 6 months, and pay another \$3000 at
the end of the year.  Alternatively, the client could sign the lease and
pay \$6000 a year later.  Since there is a legal action sequence in which 
the client makes a 6-month payment and one in which the client does not, 
we say that the client is permitted, but not obligated to make the payment.
On the other hand, if the client doesn't make the 6-month payment, then 
the only legal sequence she can be following is the second one.  Therefore,
she is obligated to complete that sequence by paying \$6000 at the end of 
the year.
}
To illustrate our notions of permission and obligation, consider the mortgage
example in which the client must pay either \$1500 between the first and fourth 
or \$1525 between the fourth and the fourteenth of every month from 1/1/01 to 
1/1/02.  For the first month, there are two legal action sequences.  The client 
could pay \$1500 before the fourth.  Alternatively, 
the client could pay \$1525 between the fourth and the fourteenth.  Since there 
is a legal action sequence in which the client pays before the fourth and one in 
which the client does not, we say that the client is permitted, but not obligated, 
to make the earlier payment.  If the client doesn't make the earlier payment, then 
the only legal sequence she can be following is the second one.  In this case, she 
is obligated to complete that sequence by paying \$1525 before the fourteenth.  

Why are we designing {\it a logic} for reasoning about licenses?
A logic provides us with a formal language in which to write
properties and specifications.
In addition, 
it allows us to check in a provably correct way that a
property or specification holds for a particular set of licenses and,
in the case of specification, a client's behavior. 
We can automate the analysis, by developing model checking techniques.
It turns out that standard model checking procedures 
(as given in \cite{Clarke99}) apply to our framework.  These procedures   
can form the foundation of enforcement mechanisms that are well-grounded in 
formal methods. 

The design 
of our logic
was strongly influenced by the work of Halpern and van der
Meyden \citeyear{Halpern01,Halpern01a} on reasoning about SPKI/SDSI.  
It is also reminiscent of deontic logic approaches, which aim at 
reasoning about ideal and actual behavior \cite{Meyer93}.  Deontic 
logic has been used extensively to analyze the structure of normative 
law and normative reasoning in law. (For examples, please see \cite{Wieringa93} 
and the references therein.)  
\commentout{
For ease of exposition, we assume licenses are 
written in a very simple, regular language and that the application has only 
one client and one provider.  Our framework can be modified in a 
straightforward manner to reason about different license languages.  It is 
also easy to extend the logic to multiple clients and providers. 
} 
In the next section, we introduce our logic.
Section~\ref{s:verification} examines the 
complexity of
checking that
a license property or client behavior specification holds. 
In Section~\ref{s:gunter}, we show that our logic can be
adapted to different license languages, by replacing our regular
language with a variant of \emph{DigitalRights} \cite{Gunter01}. We
discuss related work in Section~\ref{s:related}. Proofs of our
technical results can be found in the appendix.

\section{The logic}\label{s:logic}
We want to reason about licenses and client's actions with respect to 
licenses.  
To do this, we
introduce a logic, $\llic$,  that allows us to talk about licenses and actions.  
Formulas in $\llic$ include permission and obligation operators, as well as temporal 
operators, because we want to write formulas that represent interesting properties and
specifications; the ones that state the conditions under which actions are permitted
or obligatory.   
In this section, we give the syntax for
our logic, followed by its semantics.

\newcommand{\Circ}{\mbox{{\small $\bigcirc$}}}
\renewcommand{\cP}{\wp}
\newcommand{\Names}{\mathit{Names}}
\newcommand{\Devices}{\mathit{Devices}}
\newcommand{\Act}{\mathit{Act}}
\newcommand{\Works}{\mathit{Works}}
\newcommand{\Lic}{\mathit{Lic}}
\newcommand{\fnL}[1]{\mathcal{L}\intension{#1}}
\newcommand{\fnM}[1]{\mathcal{M}\intension{#1}}
\newcommand{\fnI}[1]{\mathcal{I}\intension{#1}}
\newcommand{\fnA}[1]{\mathcal{A}\intension{#1}}
\newcommand{\fnN}[1]{\mathcal{N}\intension{#1}}

\subsection{Syntax}
The syntax of $\llic$ has three
categories; formulas ($\phi,
\psi, \ldots$), actions ($\alpha, \ldots$), and licenses
($\ell, \ldots$).  Their definitions assume a set 
$\Names$ of license names, a set  $\Works$ of works
(i.e. resources), 
and a set  $\Devices$ of devices
(i.e. ways to access resources).
Actions are taken from a set 
$\Act = \{\mathsf{render}[w,d] \sep
w\in\Works,d\in\Devices\}\cup\{\mathsf{pay}[x]\sep x\in\R\}\cup\{\bot\}$,
where $\bot$ represents the null or ``do nothing'' action. 
(For simplicity, we consider only render and pay actions, as was
done in \cite{Gunter01}.)  
Also, we let $\Lic$ be the set of licenses $\ell$.
In the following formal description, $n\in\Names$ and $a\in\Act$.
\ifCorr
\begin{eqnarray*}
\phi & ::= & n:\ell ~|~ \alpha ~|~ P\alpha ~|~ 
\phi_1\wedge\phi_2 ~|~ \neg\phi ~|~ \Circ\phi ~|~  \Box\phi ~|~
\phi_1\cU\phi_2\\
\alpha & ::= & (a,n) ~|~ (\overline{a},n)\\
\ell & ::= & a ~|~ \ell_1~\ell_2 ~|~ \ell^* ~|~ \ell_1\cup\ell_2
\end{eqnarray*}
\else
\begin{eqnarray*}
\phi & ::= & n:\ell ~|~ \alpha ~|~ P\alpha ~|~ 
\phi_1\wedge\phi_2 ~|~ \neg\phi ~|~ \Circ\phi ~|~ \\ & & \Box\phi ~|~
\phi_1\cU\phi_2\\
\alpha & ::= & (a,n) ~|~ (\overline{a},n)\\
\ell & ::= & a ~|~ \ell_1~\ell_2 ~|~ \ell^* ~|~ \ell_1\cup\ell_2
\end{eqnarray*}
\fi
Intuitively, $n:\ell$ means ``the license whose legitimate action 
sequences are described by the regular expression $\ell$ is being issued
now and will be referred to by the name $n$.''  
The primitive action $(a,n)$ means ``action $a$ is performed with
respect to license named $n$''. 
The action $(\overline{a},n)$ represents any 
action-name pair where the action is not $a$, but the 
license
name is $n$.  
$P\alpha$ 
indicates that the action expression $\alpha$ 
is permitted.
The set of formulas are closed
under $\wedge$, $\neg$, $\Box$, $\Circ$ and $\cU$, which are
well-known
operators from classical and temporal logic \cite{Goldblatt92}.%
\footnote{Recall that $\Box\phi$ means ``$\phi$ holds now and at all future times'',
$\Circ\phi$ means ``$\phi$ holds at the next time'', and $\phi_1\cU\phi_2$ means
``$\phi_2$ eventually holds and, until it does, $\phi_1$ holds''.} 
We use the 
standard
abbreviations $\phi\vee\psi$ for $\neg(\neg\phi\wedge\neg\psi)$, $\phi\rimp\psi$ for 
$\neg\phi\vee\psi$, and $\Diamond\phi$ for $\neg\Box\neg\phi$.  Also, we abbreviate 
the action $(a,n)$ as $a_n$. For instance,
$(\mathsf{render}[w,d],n)$ is written $\mathsf{render}_n[w,d]$, and
$(\bot,n)$ is written $\bot_n$. 

We use the abbreviation $O(a,n)$ to stand for $\neg
P(\overline{a},n)$. As we shall see later, the interpretation of
$O(a,n)$ is that the client is obligated to perform action $a$ with
respect to the license named $n$. 

To illustrate how our logic can be used in practice, consider the following scenario.  Suppose
an owner of an on-line journal requires a fee to be paid before each access.  This license $\ell$ can 
be written in our logic as:  
$$\ell = ((\mathsf{pay}[\mbox{fee}](\bot)^*\mathsf{render}[\mbox{journal}, d]) \cup \bot)^*,$$
where $d$ is the device that the client uses to access the journal.
Assuming the license is labeled $n$, the property that the client is not obligated to 
access the journal immediately after paying the fee can be written as:
$$\mathsf{pay}_n[\mbox{fee}] \Rightarrow \Circ (\neg O\mathsf{render}_n[\mbox{journal}, d]).$$
The specification that the client doesn't violate the license can be written 
as the family of formulas:
\[n:\ell \Rightarrow \Box[(\alpha \Rightarrow (P\alpha)) \wedge ((O\alpha) \Rightarrow \alpha)],\]
where  
$\alpha \in \{\mathsf{pay}_n[\mbox{fee}], \mathsf{render}_n[\mbox{journal}, d], \bot_n\}$.
In other words, the client only does legitimate actions and does every action that is 
required by the license once it is issued.  As a final example, we can write that, during one time 
period, the client pays \$1500 on the mortgage $m$, but doesn't pay the journal fee as: 
$$\mathsf{pay}_m[1500] \wedge \overline{\mathsf{pay}_n[\mbox{fee}]}.$$  

\commentout{
The semantics of the logic is based on the notion of a world. A world
includes the licenses that have been issued, as well as the actions
performed by the client, be it payment actions or rendering
actions. Since time is important, a world is indexed by time units. We
assume that time is discrete and in fact representable using
integers. A \emph{world} (or a \emph{run}) is a function
$r:\N\longrightarrow \cP(\Names\times\cL)\times\cP(\Names\times\Act)$, where
at every time instant $t$, $r(t)=(NL,NA)$ with $NL\subseteq
\Names\times\cL$ is the set of pairs of name-license issued at time $t$,
and $NA\subseteq \Names\times\Act$ is the set of pairs of name-action
performed by the client at time $t$.  Although more than one action
can occur at any given time, we impose the restriction that at most
one action \emph{per license} can occur. For simplicity, when
$r(t)=(NL,NA)$, we write $\mathit{lic}(r,t)=NL$ and
$\mathit{act}(r,t)=NA$.  }

\subsection{Semantics}

To formalize the intuitions given above, we base our semantics on the
notion of a run. When defining a run, we make the standard assumption
that time is discrete and can, in fact, be represented using
nonnegative integers.  A run $r$ associates each time $t$ with a pair
$(L,A)$, where $L$ is the set of named licenses issued at that time
(a named license is a pair $(n,\ell)$ of a name $n$ and a license
$\ell$),
and $A$ is a function giving, for each license name $n$, an action $A(n)$
performed by the client at that time (or $\bot$ if no action was
performed with respect to $n$). Formally, a run is a function
$r:\N\longrightarrow\cP(\Names\times\Lic)\times\Act^{\Names}$
 such that no name is paired with more than one license throughout
the entire run.
Recall that $\Act^{\Names}$ is the set of all functions from $\Names$ to
$\Act$.  
Our
approach imposes the restriction that, at
most, one action per time \emph{per named license} can occur.  We do not
need this limitation, but it simplifies the exposition. In essence, we
are trading the ability to handle the class of licenses where a client
must do multiple actions simultaneously for a simple definition of a
license where concurrent actions are not handled. For notational
convenience, given a run $r$ and time $t$ with $r(t)=(L,A)$, we define 
$\mathit{lic}(r,t)$ to be the set of named licenses issued in run $r$ at 
time $t$, that is, $\mathit{lic}(r,t)=L$; similarly,
we define $\mathit{act}(r,t)$ to be the set of 
action and license name pairs performed in run $r$ at time $t$, that is,
$\mathit{act}(r,t)=\{(A(n),n) \sep n\in\Names\}$.  
Finally, we say that a license $(n:\ell)$ is {\it active} at time $t$ in 
run $r$ if there exists a time $t' \le t$ such that $(n:\ell) \in \mathit{lic}(r,t')$  

While a run captures 
the client's actions,
an interpretation states what is
permitted.
Formally, a
permission interpretation $P$ is a function
$P:\N\longrightarrow\cP(\Act\times\Names)$ that is used to give a
meaning to permissions.  Intuitively, if $(a,n)\in P(t)$ then at time
$t$, the client is permitted to perform action $a$ 
with respect to
license name $n$.  In other words, the client is allowed to do
an $a_n$ action. 
\commentout{
We define the function $\intension{\alpha}_A$ to convert an action
expression $\alpha$ into a set of name-action pairs as follows:
\begin{eqnarray*}
\intension{(a,n)}_A & = & \{(a,n)\}\\
\intension{\alpha_1\wedge\alpha_2}_A & = & \intension{\alpha_1}_A\inter\intension{\alpha_2}_A\\
\intension{\neg\alpha}_A & = & (\Act\times\Names)-\intension{\alpha}_A
\end{eqnarray*}
} 
We 
want 
the 
interpretation of permissions  to
match the permissions implied 
by the run.
To define this requirement formally, we first give a mapping 
that relates licenses to action sequences.
We then use this mapping to 
find the permission interpretation that permits an action if and only if
the run implies the permission.

Following the lead of Gunter \emph{et al.} \citeyear{Gunter01},
we associate each license with a set of 
traces.  
\commentout{
In our discussion, a \emph{trace} refers to a sequence of
actions;\footnote{Gunter \emph{et al.} \citeyear{Gunter01} use the
term \emph{reality} for this concept, although their formal definition
is different.}  the notation $s_1\cdot s_2$ denotes the concatenation
of 
two action sequences 
$s_1$ and $s_2$; and a trace $s_1$ is
said to be a
\emph{prefix} of trace $s_2$ if there is some trace $s$ such that
$s_1 \cdot s = s_2$.  Given a license $\ell$, we
construct 
a function $\fnL{\ell}$ by induction on the structure of $\ell$:
}
In our discussion, a \emph{trace} refers to a sequence of
actions.\footnote{Gunter \emph{et al.} use the
term \emph{reality} for this concept, although their formal definition
is different.}  The notation $s_1\cdot s_2$ denotes the concatenation
of two sequences of actions $s_1$ and $s_2$ where $s_1\cdot s_2 = s_1$
if $s_1$ is infinite.  A trace $s_1$ is said to be a
\emph{prefix} of trace $s_2$ if there is some trace $s$ such that
$s_1 \cdot s = s_2$.  

We construct a function $\fnL{\ell}$ by induction on the structure of 
a given license $\ell$:
\begin{eqnarray*}
\fnL{a} & = & \{a\}\\
\fnL{\ell_1~\ell_2} & = 
& \{s_1\cdot s_2 \sep s_1\in\fnL{\ell_1} \mbox{ and } s_2\in\fnL{\ell_2}\}\\
\fnL{\ell_1\cup\ell_2} & = &
\fnL{\ell_1}\cup\fnL{\ell_2}\\
\fnL{\ell^*} & = & 
\bigcup_{n\geq 0} \{s_1\cdot\ldots\cdot s_n \sep s_i\in\fnL{\ell}\}.
\end{eqnarray*}
The function $\fnL{\ell}$ gives the set of 
traces allowed by the license. We define the function
$\fnI{\ell}$ to provide the infinitary version of the
sequences corresponding to $\ell$, by essentially appending infinitely 
many $\bot$ actions at the end of each sequence.
Formally, $\fnI{\ell} = \{s\cdot\bot^\infty \sep s\in\fnL{\ell}\}$.
Finally, a sequence of action $s$ is said to be \emph{viable} for $\ell$ if $s$ is a prefix of some trace in $\fnI{\ell}$. 

We are now ready to define the interpretation $P_r$ corresponding to
run $r$.  Given a named license $(n,\ell)$ issued
at time $t_1$ in a run $r$, the \emph{action-sequence} of $n$
up to time $t_2$, denoted 
$r[n,t_2]$,
is the sequence
$a_0a_1\cdots a_{t_2-t_1-1}$ such that:
\[ a_i = \left\{ \begin{array}{ll}
   a & \mbox{if $(a,n)\in\mathit{act}(r,t_1+i)$}\\
   \bot & \mbox{otherwise.}
                 \end{array}\right. \]
Since we restricted a run to only allow one action per license per
time unit, the notion of an action-sequence is well-defined.  The
interpretation $P_r$ corresponding to a run $r$ is defined as
follows. For all times $t\geq 0$, $P_r(t)$ is the smallest set such that for all
license names $n\in\Names$ and  actions $a\in\Act$, $(\bot,n)\in P(t)$ 
if the license $(n,\ell)$ is not active and $(a, n)\in P(t)$
if the license is active and $r[n,t] \cdot a$ is viable for $\ell$.  
\commentout{ 
\begin{itemize}
\item if a license named $n$ is never issued in a run, then
$(\bot,n)\in P(t)$ for all $t\geq 0$.  And 
\item if a license $(n,\ell)$ is issued at time $t_0$, then 
$(\bot,n)\in P(t)$ for all times $t<t_0$ and,
if $r[n,t] \cdot a$ is viable for $\ell$, then $(a, n)\in
P(t)$, for all times $t\geq t_0$ and all actions $a\in\Act$.
\end{itemize}
}%
\commentout{
We are interested in interpretations that enforce exactly the
permissions
implied by the run. 
To obtain one, define an ordering on interpretations such that
$P\leq P'$ if, for all times $t$, $P(t)\subseteq P'(t)$.  Thus, 
$P\leq P'$ if for all times $t$, every action permitted by $P$ is also
permitted by $P'$. 

It is easy to see that this ordering gives the set of interpretations
the structure of a lattice.
We say that an element $P$
in a set of interpretations $S$
is minimal in $S$ if $P\leq P'$ for all $P'\in S$.
Let $\mathcal{P}_r$ be the set of interpretations consistent with run $r$. 
It is easy to verify that the interpretation $P_r$ defined by:
\[ P_r(t) = \bigcap_{P\in\mathcal{P}_r} P(t) \]
is the unique minimal interpretation in $\mathcal{P}_r$. We will use
$P_r$ as the interpretation for permissions in run $r$. 
}  %

To understand the meaning of an action expression, $\alpha$, we need a way 
to associate it with name-action pairs.  We do this by defining
a mapping $\fnA{\alpha}$ from expressions to sets of pairs.  Clearly, an 
action expression $(a, n)$ should be mapped to the pair $(a, n)$.  
The complement action $(\overline{a},n)$ is mapped to the set of
actions different from $a$, but associated with the same license name
$n$. Formally,
\begin{eqnarray*}
\fnA{(a,n)} & = & \{(a,n)\}\\
\fnA{(\overline{a},n)} & = & \{ (b,n) ~|~ b\not=a\}.
\end{eqnarray*}
\commentout{
A natural alternative
is
to associate a complementary action
$\overline{(a,n)}$ with the set of actions that are not associated
with the license named $n$. 
}
Contrary to intuition, we do not associate the complement of a
name-action pair with  the largest set of name action pairs that does
not include it.   This mapping 
has
unfortunate consequences, because it ignores 
the intuitive independence between licenses.  For example, it 
allows us 
to deduce that the client can do any action with respect to any license other
than the mortgage, if the client is permitted to not make a mortgage payment.  
Statements concerning one set of licenses should not 
be used to deduce anything about any other license.  
As an example
of our approach, 
recall the situation in which 
the client pays \$1500 on the mortgage, but doesn't pay the journal
fee.  The action expressions $\alpha_1$ and $\alpha_2$ used to express 
these actions are $\mathsf{pay}_m[1500]$ and
$\neg\mathsf{pay}_n[\mbox{fee}]$, respectively. 
Applying the above definition, 
$\fnA{\alpha_1} = \{(\mathsf{pay} [1500], m)\}$, and 
$\fnA{\alpha_2}= \{(a, n)  \sep  a \ne \mathsf{pay}[\mbox{fee}]\}$.
 Hence, the
actions $\alpha_1$ and $\alpha_2$ mean that 
``the client is
paying \$1500 with respect  to $m$ and doing some action other than
paying the fee with respect to $n$''. 

We now define what it means for a formula $\phi$ to be true
(or satisfied) at a run $r$ at time $t$, 
written $r,t\sat\phi$, by induction on the structure of
$\phi$:
\begin{itemize}
\item[] $r,t\sat n:\ell~~$ if $(n,\ell)\in \mathit{lic}(r,t)$,
\item[] $r,t\sat \alpha~~$ if
$\exists(a,n)\in\fnA{\alpha}$ s.t. $(a,n)\in\mathit{act}(r,t)$,
\item[] $r,t\sat P\alpha~~$ if
$\exists(a,n)\in\fnA{\alpha}$ s.t. $(a,n)\in P_r(t)$,
\item[] $r,t\sat \Circ\phi~~$ if $r,t+1\sat\phi$,
\item[] $r,t\sat \Box\phi~~$ if for all $t'\geq t$, $r,t'\sat\phi$,
\ifCorr
\item[] $r,t\sat \phi\cU\psi~~$ if $\exists t'\geq t$ s.t.
$r,t'\sat\psi$ and $r,t''\sat\phi$ for all $t''$ with $t'>t''
\geq t$,
\else
\item[] $r,t\sat \phi\cU\psi~~$ if $\exists t'\geq t$ s.t.
$r,t'\sat\psi$ and\\ $~\qquad\qquad\qquad r,t''\sat\phi$ for all $t''$ with $t'>t''
\geq t$,
\fi
\item[] $r,t\sat \neg\phi~~$ if $r,t\not\sat\phi$,
\item[] $r,t\sat \phi\wedge\psi~~$ if $r,t\sat\phi$ and
$r,t\sat\psi$. 
\end{itemize}

If a formula $\phi$ is true at all times in a run $r$, we say $\phi$
is \emph{valid} in 
$r$
and write 
$r\sat\phi$. If $\phi$ is valid in all runs $r$, we
simply say $\phi$ is valid
and write 
$\sat\phi$. 
\footnote{In an earlier version of this paper \cite{Pucella02}, we considered two
related semantics for formulas, in the spirit of the logics presented
by Halpern and van der Meyden \citeyear{Halpern01,Halpern01a}. The
first semantics, called the open semantics, was defined with respect
to an arbitrary interpretation $P$. The second semantics,
called the closed semantics, was defined from the open semantics by
taking the minimal interpretation, as we do in this
paper. Intuitively, the closed semantics assumes that the run contains
all the information relevant to interpret the formulas. This is often
referred to as the \emph{closed-world assumption}. In other words, 
if a permission is not implied by the run, then it is not permitted.
In contrast, the open semantics admits that the run may not encode all
the information, and therefore one cannot infer that an action is not
permitted simply because it is not implied by the run.}

Various properties of permission ($P$) and obligation ($\neg P(\overline{a}, n)$)
follow from the above semantics.
In particular, we can see that $O(a,n)$ is true in a run $r$ at time
$t$ if and only if $(a,n)$  is the only action-name pair in $P_r(t)$.
In other words, an action is obligated if and only if it is the only
permitted action. This is a consequence of the following proposition:
\pro\label{p:validp}
For all action expressions $(a,n)$, the formula $P(a,n)\vee
P(\overline{a},n)$  is valid.
\epro
Hence, if $P(\overline{a},n)$ is not true at a point, $P(a,n)$ must be 
true. Another consequence of the above proposition is that
$O(a,n)\rimp P(a,n)$ is  valid. 
These properties show that our operators $P$ and $O$, although defined 
exclusively from the traces of the licenses issues in a run, satisfy
some of the classical properties of deontic logic operators, as given
for instance in \cite{Follesdal81}. 
These properties are 
a consequence of 
our prescribed semantics
and, as such, suggest
a certain deontic interpretation. In particular, the validity of 
$O(a,n)\rimp P(a,n)$ indicates that
obligation should be read as ``must'' and not as ``ought''.
It also 
reflects the fact that we cannot express conflicting prohibitions and
obligations in our framework.

\subsection{Encoding finite runs and licenses}\label{s:encodings}

In this section, we show that any run can be ``encoded'' as a formula in our logic,
provided that the run is finite.  By finite, we intuitively mean that nothing 
happens after a given time, and each time instant, only finitely many
licenses are issued and non-$\bot$ actions are performed. Formally, a
run $r$ is finite if there exists a natural number $t_f$ such that :
\begin{itemize}
\item for all $t\leq t_f$, $\mathit{lic}(r,t)$ is finite,
\item for all $t\leq t_f$, $\{n ~:~ (a,n)\in\mathit{act}(r,t),
a\not=\bot\}$ is finite,
\item for all $t>t_f$, $\mathit{lic}(r,t)=\emptyset$, and
\item for all $t>t_f$, $(a,n)\in\mathit{act}(r,t)$ implies $a=\bot$.
\end{itemize}

For convenience, we write $\Circ^k\phi$ for the formula
$\Circ\cdots\Circ\phi$
that has k occurences of the $\Circ$ operator before $\phi$.  
Given a finite run $r$, define $N_r$ to be the set of
license names issued in $r$.
Formally, $N_r=\{n ~:~ \exists t,\ell.(n,\ell)\in\mathit{lic}(r,t)\}$.
Define
\[ \psi_r = \psi_0\wedge\Circ\psi_1\wedge\Circ^2\psi_2\wedge\cdots\wedge\Circ^{t_f}\psi_{t_f}\wedge\Circ^{t_f+1}\Box\psi_e,\]
where $t_f$ is 
the last time ``something happened'' in the run, 
$\psi_e$ is $\bigwedge_{n\in
N_r}(\bot,n)$,  and $\psi_t$,  
which encodes the state of the run at time $t$, is:
\[\psi_t = \bigwedge_{(a,n)\in\mathit{act}(r,t) \atop n\in
N_r}(a,n)\wedge\bigwedge_{(n,\ell)\in\mathit{lic}(r,t)}n:\ell.\] 
Finally, let $N_\phi$ be the set of license names appearing in formula $\phi$, 
defined in the obvious way.  The  following proposition formalizes the fact 
that $\psi_r$ captures the important aspects of the run $r$.  
\pro\label{p:runencoding}
If $r$ is a finite run and $N_\phi\subseteq N_r$, 
then $r,t \sat \phi$ iff $\sat\psi_r\rimp \Circ^t\phi$.
\epro

It is interesting to note 
that $\psi_r$ does not specify explicitly the
permissions implied by the run. Intuitively, this is because the
information encoded in $\psi_r$ is sufficient for the permissions to
be uniquely determined. 
To formalize this intuition, we show the more general result that 
issuing a license results in the client's actions implying a particular
set of permissions.  

We use some notation from the theory of regular languages to formalize
the general result.  Specifically, we let 
$\epsilon$ represent the empty
action sequence
and we extend the set of licenses to include $0$ and $1$ where 
$\fnL{0}=\emptyset$ and $\fnL{1}=\{\epsilon\}$.  We also define complementary
functions $S(\ell)$ and $D_a(\ell)$ where $\ell$ is a regular
expression.  For any
action sequence $a_0, a_1, \dots, a_n\in\fnL{\ell}$, $S(\ell)$ is the set of actions
containing $a_0$ and $D_{a_0}(\ell)$ is a regular expression such that 
$a_1, \dots, a_n\in\fnL{D_{a_0}(\ell)}$.
Formally,
$S(0)=\emptyset$, $S(1)=\emptyset$, $S(a) =  \{a\}$,
$S(\ell_1\ell_2) = S(\ell_1)$ if  $\epsilon\not\in\fnL{\ell_1}$ and
$S(\ell_1)\cup S(\ell_2)$ otherwise,  $S(\ell_1\cup\ell_2) =
S(\ell_1)\cup S(\ell_2)$, and $S(\ell^*) = S(\ell)$.  $D_a(\ell)$ is called the 
Brzozowski derivative of $\ell$ with respect to $a$ \cite{Brzozowski64}.  Its formal definition 
is: $D_a(a) = 1$, $D_a(b) = 0$,
$D_a(\ell_1\ell_2) = D_a(\ell_1)\ell_2$ if $\epsilon\not\in\fnL{\ell_1}$ and
$(D_a(\ell_1)\ell_2)\cup(D_a(\ell_2))$ otherwise,
$D_a(\ell_1\cup\ell_2)=D_a(\ell_1)\cup D_a(\ell_2)$, and $D_a(\ell^*)
= D_a(\ell)\ell$. 

Given these definitions, we inductively define a family of formulas
for each named license $(n,\ell)$.  For any action sequence 
$a_0 a_1 \cdots a_n\in\fnL{\ell}$, the formulas say that $a_0$ 
is permitted and if the client does the action sequence $a_0\cdots a_{i-1}$, 
then the client is permitted to do $a_i$ in $i$ time steps.  Formally: 
\begin{eqnarray*}
\phi_{n,\ell}^0 & = & \bigwedge_{a\in S(\ell)} P(a,n)\\
\phi_{n,\ell}^{i+1} & = & \bigwedge_{a\in S(\ell)} \left( P(a,n)
\wedge \left( (a,n)\rimp \Circ \phi_{n,D_a(\ell)}^i\right) \right).
\end{eqnarray*}
The following proposition formalizes 
the intuition that by issuing a license, we force the client's actions
to imply a particular set of permissions.  
\pro\label{p:licencoding}
For any license $\ell$, the formulas $n:\ell\rimp\phi_{n,\ell}^i$ are
valid, for $i=0,1,2,\ldots$. 
\epro
Hence, if the formula $\psi_r$ represents the finite run $r$ in the
sense of Proposition~\ref{p:runencoding}, 
then
every named license
$(n,\ell)$ issued in run $r$ will imply the formulas
$\phi_{n,\ell}^i$, as per Proposition~\ref{p:licencoding}. 
Because the conjunction of the actions specified in $\psi_r$ and the formula
$\phi_{n,\ell}^i$ implies the permissions that hold for run $r$ for 
$i$ time steps, Proposition~\ref{p:runencoding} is true even though $\psi_r$
does not specify permissions explicitly.   

\section{Satisfiability and verification}\label{s:verification}

In this section, we 
examine
the complexity of reasoning using $\llic$ and 
discuss
a technique for automatically checking if a client behavior
specification is satisfied in a given run. 
As we mentionned in the introduction, we are fundamentally interested
in two classes of questions
does
a set of licenses 
have
certain properties
and does 
a client's actions with respect to a set of licenses meet
particular specifications. 
The first question can be rephrased as ``does a set of licenses imply a property, 
regardless of what the client does, which licences are issued, and when the licenses
are issued?''.  In other words, the first question corresponds to asking if a 
formula in our logic is valid (i.e., true in all runs).  The second question can be 
rephrased as ``does a specification hold for a given sequence of client actions and 
licenses issued?''  In other words, the second question corresponds to asking if a 
formula in our logic is true in a given run.  

To answer the first question, we investigate the complexity of
our satisfiability problem (i.e. the problem of determining for any given
$\llic$ formula $\phi$ if there exists a run $r$ and a time $t$ such that
$r,t\sat\phi$).  
We can reduce the satisfiability problem for our logic
to the satisfiability problem for a ``simpler'' logic, Linear Temporal 
Logic (LTL),
which is well-known in the formal verification community. 
LTL is essentially 
a propositional logic with temporal operators. 
To distinguish the LTL operators from the temporal operators
in $\llic$, 
we 
use CTL syntax for LTL.  Specifically, an LTL formula $F$ is defined as:
\[F  ::= p ~|~ F_1\wedge F_2 ~|~ \neg F ~|~ \mathbf{X}F ~|~
\mathbf{G}F ~|~ F_1\mathbf{U}F_2\]
where 
$p$ is a primitive proposition, 
$\mathbf{X}F$ means that $F$ holds at the next time, $\mathbf{G}F$ means that $F$ holds now 
and at all future times, and $F_1\mathbf{U}F_2$ means that $F_2$ eventually holds and, until
it does, $F_1$ holds.  
Models for LTL are 
linear structures 
of the form
$M=(S,L)$, where $S=\{s_0,s_1,s_2,\ldots\}$
is a set of states 
and $L$ assigns to every state 
in
$S$ the primitive propositions that are true in that state. 
The definition of the 
satisfiability of an LTL formula $F$ in a linear structure $M$ at state $s$,
written $M,s\sat_L F$, is straightforward. We refer to \cite{Clarke99} 
for more detail. 
The key property of LTL that we will use is that the satisfiability
problem for LTL is PSPACE-complete \cite{Sistla85}.

It is straightforward to encode a formula $F$ in LTL as a formula
$\phi$ in $\llic$ in such a way that $F$ is satisfiable if and only
if $\phi$ is satisfiable.
Therefore, the satisfiability problem for $\llic$ is PSPACE-hard.
What is more
interesting is that there is a polynomial reduction from the
satisfiability problem for $\llic$ to the satisfiability problem for
LTL. At the heart of this reduction is a way to encode our logic into
LTL.  

The first step of the reduction is to show that if a formula
$\phi$ is satisfiable in $\llic$, then it can be translated into a
satisfiable formula $\phi^T$ in LTL. We will do this directly, by
showing that we can in fact transform the run $r$ in which $\phi$ is
true into a linear structure $M_r$ in which $\phi^T$ is true. Let
$\Phi_0$ be the set of primitive propositions that we will use in our
formula encoding, inclduing primitive propositions
$\mathsf{issued}(n,\ell)$ for every name $n$ and license $\ell$, and
$\mathsf{done}(a,n)$, $\mathsf{permitted}(a,n)$ and
$\mathsf{obligated}(a,n)$ for each action $a$ and name $n$. 

Given a run $r$, we construct a linear model $M_r = (S,L)$ where $S = 
\{s_0, s_1, s_2, \ldots\}$. For each state $s_t$, which corresponds to
the run at time $t$, $L(s_t)$ is defined as the smallest set such that:
\begin{itemize}
\item if $(n,\ell)\in\mathit{lic}(r,t)$, then
$\mathsf{issued}(n,\ell)\in L(s_t)$,
\item if $(a,n)\in\mathit{act}(r,t)$, then $\mathsf{done}(a,n)\in
L(s_t)$,
\item if $(a,n)\in P_r(t)$, then $\mathsf{permitted}(a,n)\in L(s_t)$,
\item if $(a,n)\in P_r(t)$ is the only action associated with license
name $n$ in $P_r(t)$, then 
$\mathsf{obligated}(a,n)\in L(s_t)$. 
\end{itemize}
Given this structure $M_r$, it should be clear how to translate a
$\llic$ formula $\phi$ true in $r$ into a formula $\phi^T$ true in
$M_r$. In particular, the following translation works:
\begin{itemize}
\item $(n:\ell)^T = \mathsf{issued}(n,\ell)$. 
\item  $(a,n)^T = \mathsf{done}(a,n)$ and $(\overline{a},n)^T = \neg\mathsf{done}(a,n)$. 
\item $(P(a,n))^T = \mathsf{permitted}(a,n)$ and
$(P(\overline{a},n))^T = \neg \mathsf{obligated}(a,n)$. 
\item $(\phi_1\wedge\phi_2)^T = \phi_1^T\wedge\phi_2^T$ and $(\neg\phi)^T =
\neg\phi^T$.
\item  $(\Circ\phi)^T  = \mathbf{X}\phi^T$, $(\Box\phi)^T =
\mathbf{G}\phi^T$, and $(\phi_1\cU\phi_2)^T =
\phi_1^T\mathbf{U}\phi_2^T$. 
\end{itemize}
It is straightforward to see that the above translations preserve the
truth of the formula. In fact, something stronger holds, which will be
useful later in this section: 
\pro\label{p:ltlmodelc}
$r,t\sat \phi$ iff $M_r,s_t \sat_L \phi^T$.
\epro
This means that if $\phi$ is satisfiable in our logic,
then $\phi^T$ is satisfiable in LTL. However, the converse does not
hold. In particular, $\phi^T$ may be satisfiable in an LTL structure
that does not correspond to any run. We somehow need a way to restrict
the LTL structures considered, to ensure that they correspond to runs
in $\llic$. Intuitively,
we need to account in LTL for the notions that are implicit in the
$\llic$ semantics.  In particular, 
we must enforce our requirements that two actions are never done for
the same license at the same time, two licenses are never labeled with
the same name, an obligation implies exactly one action is permitted
for the license, a 
client is only permitted to do actions other than $\bot$ for active
licenses, and issuing a license implies various facts as discussed in
Section~\ref{s:encodings}.  It is easy to state all but the last of
these in LTL. 

Since we will only need to satisfy the above restrictions as they
pertain to a given formula $\phi$, we enforce those restrictions over
the actions, license names, and licenses appearing in $\phi$. In
general, let $A$ be a finite set of actions, $N$ be a finite set of
license names, and $L$ be a finite set of named licenses. The
restriction that at most one action is done per license name per time
is expressed by the following LTL formula $\mathsf{Done}_{A,N}$: 
\[\mathbf{G}\bigwedge_{a\in A \atop n\in N} 
  \left(\mathsf{done}(a, n) \Rightarrow \bigwedge_{a' \in A\atop a'\not= a}
  \neg\left(\mathsf{done}(a', n)\right)\right).\] 
The restriction that a license name in $N$ is never associated with
more than one license in $L$ is expressed by the LTL formula
$\mathsf{Issued}_{L}$: 
\[\mathbf{G}\bigwedge_{(n,\ell)\in L}
  \left(\mathsf{issued}(n,\ell) \Rightarrow 
   \bigwedge_{(n',\ell')\in L\atop n'=n}\mathbf{G}\neg(\mathsf{issued}(n',\ell'))\right).\]
The restriction that obligation is an abbreviation for only being
allowed to do one action with respect to a license is expressed by the 
LTL formula $\mathsf{Obl}_{A,N}$: 
\[\mathbf{G}\bigwedge_{a\in A\atop n \in N} 
   \left(\begin{array}{l}
    \mathsf{obligated}(a, n) \riff \\
    \quad\left(\begin{array}{l}
                \mathsf{permitted}(a,n)\wedge \\
                \bigwedge\limits_{a' \in A\atop a'\not=a}
                   \neg(\mathsf{permitted}(a', n)))\end{array}\right)
         \end{array}\right).\] 
The restriction that a client can only do $\bot$ actions with respect
to an unissued license is expressed by the LTL formula
$\mathsf{Unissued}_{L}$:
\[\bigwedge_{(n,\ell)\in L}
\left(\mathsf{obligated}(\bot, n) ~\mathbf{U} ~\mathsf{issued}(n,\ell)\right).\]

To state the consequences of issuing a named license $(n, \ell)$, we
first construct a nondeterministic finite automaton (NFA) that accepts
the same language as $\ell$ (when $\ell$ is viewed as a regular
expression), and encode the transition relation of the automaton as an
LTL formula. Formally, we construct the $\epsilon$-free NFA
representing $\ell$ as $A_{n} = (Q_n, \Delta_n, S_n, F_n)$ where $Q_n$
is the set of states, $\Delta_n$ is the transition function, $S_n$ are
the start states, and $F_n$ are the final states. For convenience, we
will write $\Delta_n(q)$ for $\{a ~:~ \exists q'\in Q_n.(q,a,q')\in
\Delta_n\}$ and $\Delta_n(q,a)$ for $\{q' ~:~
(q,a,q')\in\Delta_n\}$. We assume that we have primitive propositions
in $\Phi_0$ to represent the states of the automaton, namely
$\mathsf{instate}(n,q)$ for all $q\in Q_n$, and a primitive
proposition $\mathsf{over}(n)$ to represent the fact that we have
stopped taking transitions in the automaton (for instance, because the
client performed an action that was not permitted). The ``effect'' of
taking a transition (from a finite set $A$ of actions) in a state $q$
of $A_n$ can be represented by the following LTL formula
$\mathsf{Trans}_{A,q}$:  
\[\begin{array}{l}
\mathsf{instate}(n,q)\rimp\\
\quad\left(\begin{array}{l}
\bigwedge\limits_{a\in\Delta_n(q)}(\mathsf{permitted}(a,n))\wedge \\
\bigwedge\limits_{a\in\Delta_n(q)}\left(\begin{array}{l}
                                   \mathsf{done}(a,n)\rimp\\
                                   \quad\bigvee\limits_{q'\in\Delta(q,a)}\mathbf{X}(\mathsf{instate}(n,q'))\end{array}\right)\wedge\\
\bigwedge\limits_{a\in A \atop
a\not\in\Delta_n(q)}(\neg\mathsf{permitted}(a,n))\wedge\\
\bigwedge\limits_{a\in A \atop
a\not\in\Delta_n(q)}\mathsf{done}(a,n)\rimp\mathbf{X}(\mathsf{over}(n))\end{array}\right).\end{array}\] 

We also need a statement to the effect that the automaton $A_{n}$
can only be in one state at any given time, or in a state satisfying
$\mathsf{over}$. This is expressed by the following LTL formula $\mathsf{States}$:
\[\begin{array}{l}
\left(\mathsf{over}(n)\rimp\bigwedge\limits_{q\in
Q_n}\neg\mathsf{instate}(n,q)\right)\wedge\\
\bigwedge\limits_{q\in Q_n} \left(\begin{array}{l}\mathsf{instate}(n,q)\rimp\\
\quad\left(\neg\mathsf{over}(n)\wedge\bigwedge\limits_{q'\in Q_n \atop q'\not=q}
\neg\mathsf{instate}(n,q')\right)\end{array}\right).
  \end{array}\]
The encoding of the NFA $A_{n}$ is then expressed by the
following LTL formula $\mathsf{NFA}_{n,\ell,A}$, which asserts the
initial states of the automaton, as well as encoding all the
transitions, including the transitions from the states where
$\mathsf{over}(n)$ holds:
\[ \begin{array}{l}
\left(\bigvee\limits_{q\in S_n} \mathsf{instate}(n,q)\right) \wedge \mathbf{G}(\mathsf{States})\wedge\\
\mathbf{G}\left(\begin{array}{l}\bigwedge\limits_{q\in
Q_n} \mathsf{Trans}_{A,q} \wedge\\
 (\mathsf{over}(n)\rimp
(\mathsf{obligated}(\bot,n)\wedge\mathbf{X}(\mathsf{over}(n))))\end{array}\right).\end{array}\] 

The restriction that issuing a license implies the consequences
described by the corresponding NFA is therefore expressed by the LTL
formula $\mathsf{Lic}_{L,A}$: 
\[\mathbf{G}\bigwedge_{(n,\ell)\in L} (\mathsf{issued}(n,\ell) \Rightarrow
\mathsf{NFA}_{n,\ell,A}).\]
Note that the formula corresponding to the NFA construction
guarantees that only the $\bot$ action is allowed for a completed 
license. 

We now associate with every $\llic$ formula $\phi$ the LTL formula
$\phi^I$ that captures all the implicit restrictions required for our
treatment of $\phi$. Recall from Section~\ref{s:encodings} that
$N_\phi$ represents the set of license names appearing in $\phi$. In
a similar way, define $A_\phi$ to be the set of actions explicitely
appearing in $\phi$, and define $L_\phi$ to be the set of named
licenses appearing in $\phi$ (i.e., occurrences of the $n:\ell$
formula). We take $\phi^I$ to be:
\ifCorr
\[
\mathsf{Done}_{A_\phi,N_\phi}\wedge\mathsf{Issued}_{L_\phi}\wedge\mathsf{Obl}_{A_\phi,N_\phi}\wedge\mathsf{Unissued}_{L_\phi}\wedge\mathsf{Lic}_{L_\phi,A_\phi}.\] 
\else
\[\begin{array}{l}
\mathsf{Done}_{A_\phi,N_\phi}\wedge\mathsf{Issued}_{L_\phi}\wedge\mathsf{Obl}_{A_\phi,N_\phi}\wedge\\
\quad\mathsf{Unissued}_{L_\phi}\wedge\mathsf{Lic}_{L_\phi,A_\phi}.\end{array}\]
\fi
We can formally verify that the formula $\phi^I$ does indeed capture
the implicit restrictions imposed by the semantics of $\llic$, as far
as they pertain to formula $\phi$. We can 
show: 
\pro\label{p:licmodelc}
If $M,s\sat_L\phi^T\wedge\phi^I$, then there exists a run $r$ such that 
$r,0\sat\phi$. 
\epro
Propositions~\ref{p:ltlmodelc} and \ref{p:licmodelc} can be used to
derive the following characterization of the complexity of the logic:
\thm\label{t:satisfiability}
The satisfiability problem for $\llic$ is PSPACE-complete.
\ethm
Since a formula $\phi$ is valid if and only if $\neg\phi$ is not
satisfiable, a corollary of Theorem~\ref{t:satisfiability} is that
determining if a formula $\phi$ of our logic is valid is also a
PSPACE-complete problem.
It is much easier to answer our second question.  
The above discussion in fact hints at a suitable approach:
we reduce
the model-checking problem for our logic to one for LTL and
then apply existing verification technology developed for LTL.  
More specifically, we translate the run (and associated minimial 
interpretation $P_r$) into a linear structure with a state for each time
and atomic propositions for the licenses issued, client actions, 
permissions and obligation.  
We restrict our attention to finite runs, as defined in
Section~\ref{s:encodings}, because we want to give an  
algorithm for deciding if a formula holds in a given model. (In practice, 
we expect to have a description of client behavior for a period of time 
and we want to establish permissions or obligations given that behavior; 
this can be modeled with a finite run.)  
The idea is simply to use the construction of the LTL structure $M_r$
as given earlier, and use Proposition~\ref{p:ltlmodelc}. The only
problem is that the construction of $M_r$ assumes that we have the
permission interpretation $P_r$. To construct $M_r$ efficiently, we
need a way to compute $P_r$ efficiently.
For each named license $(n,\ell)$ (finitely many by assumption), we
construct an NFA that accepts the
language represented by  $\ell$.  We associate a subset of the NFA's
states with every time $t$ after the  license is issued.
Specifically, the NFA's initial states are associated with the  
time when the license is issued.  
The states associated with any later time $t+1$ is the set of states that can be reached
by one transition from a state associated with time $t$.
For every time $t$ after the license is issued, the set of permitted actions $P_{r,n}(t)$ 
is the set of 
possible transitions from the states associated with $t$.  Finally, for 
any time $t$, $P_r(t)$ is the union of $P_{r,n}(t)$ for all licenses named $n$ issued by 
time $t$.  
This procedure constructs $P_{r,n}(t)$ in polynomial time with respect to the size of the run.   
\pro\label{p:minimalint}
There exists a polynomial time algorithm for computing the 
interpretation $P_r$ corresponding to a finite run $r$. 
\epro

Combining the computation of $P_r$ from $r$ with the construction of
the model $M_r$ given earlier
and applying known LTL model-checking
techniques, model checking can be done reasonably efficiently, at
least for 
a 
small specification $\phi$:
\thm\label{t:modelchecking}
There exists an algorithm for deciding if a formula $\phi$ is true in a 
finite run $r$ at time $t$.  Furthermore, the algorithm runs in polynomial
time with respect to the size of the model $r$ and in exponential time with
respect to the size of the formula $\phi$.
\ethm

A straightforward modification to the above procedure would allow us to check
the validity of a formula $\phi$ in a run $r$ (i.e., check that $\phi$ holds
throughout the run).  
\pro\label{p:ltlvalc}
$r\sat\phi$ iff $M_r,s_0\sat_L \mathbf{G}(\phi^T)$.
\epro

Finally, note that the model $M_r$ 
is constructed without regard to the
formula $\phi$ whose truth value we want to
check. 
Therefore, we can construct $M_r$ once and use it to model-check different 
formulas, each translated to LTL, against the run $r$.

\section{Handling different license languages}\label{s:gunter}

In discussing our logic thus far, we have assumed that the licenses
are written in a regular language.  Although a regular language has the
benefits of being well-known, simple, and fairly expressive, it is not 
difficult to imagine settings in which another license language
is more appropriate. A key feature of our logic is that it can be 
adapted in a straight-forward way to reason about licenses that are 
written in any language that has trace-based semantics.  To illustrate this 
flexibility, we will modify our logic to handle the licenses presented in Gunter 
\emph{et al.} \citeyear{Gunter01}.

\newcommand{\Gfor}{\mathsf{for}}
\newcommand{\Gupto}{\mathsf{upto}}
\newcommand{\Gpay}{\mathsf{pay}}
\newcommand{\Gupfront}{\mathsf{upfront}}
\newcommand{\Gflatrate}{\mathsf{flatrate}}
\newcommand{\Gperuse}{\mathsf{peruse}}
\newcommand{\Gon}{\mathsf{on}}

For ease of exposition, we consider a restricted version of
\emph{DigitalRights} \cite{Gunter01}.\footnote{The original
\emph{DigitalRights} allows one to specify the time at which a client can 
activate a license. Roughly speaking, we could capture this in our
model by adding license activation as an action.} The syntax of
licenses is given by 
the following grammar:
\begin{eqnarray*}
e & ::= & (\Gfor~p ~|~ \Gfor~[\Gupto]~m~p)\\
 & & \Gpay~x~(\Gupfront ~|~ \Gflatrate ~|~ \Gperuse)\\
 & & \Gfor~W~\Gon~D
\end{eqnarray*}
where $p$ is a period of time (a number of time units), $x$ is a
payment amount, $W$ is a subset of works and $D$ is a subset of
devices. 
\commentout{
Taking $H$ to be $\Gupfront$, $\Gflatrate$ or $\Gperuse$, a
license of the  form $\Gfor~p~\Gpay~x~H~\Gfor~W~\Gon~D$ intuitively
means that for   the time period indicated by $p$, the user is
required to pay $x$ (according to the schedule indicated by $H$) in
order to render any of the works in $W$  on a device in $D$. The
schedule $\Gupfront$ means that the payment of $x$ must be
made at the beginning of the time period, the schedule
$\Gflatrate$ means that the payment of $x$ must be made at the end
of the time period (irrespectively of the number of uses), and
the schedule $\Gperuse$ means that the  payment of $x$ \emph{per use}
must be made at the end of the time period. The license can be valid
for multiple time periods.  If the license starts with $\Gfor~\Gupto~m~p$, 
then the body of the license is valid for up to $m$ time periods of length
$p$
}
The terms $\Gupfront$, $\Gflatrate$ and $\Gperuse$ refer to the payment 
schedule.  The $\Gupfront$ schedule requires payment at the beginning
of the time period.  The $\Gflatrate$ and $\Gperuse$ schedules require 
payment at the end of the time period.  The difference between the two 
is that the payment for $\Gflatrate$ does not depend on the number of 
renderings, while the one for $\Gperuse$ does.  If we let $H$ be a payment
schedule ($\Gupfront$, $\Gflatrate$ or $\Gperuse$), then a license of the 
form $\Gfor~p~\Gpay~x~H~\Gfor~W~\Gon~D$ means that for the time period 
indicated by $p$, the client is required to pay $x$, according to schedule 
$H$, in order to render any of the works in $W$ on a device in $D$.  
Instead of beginning with $\Gfor~p~$, a license can start with $\Gfor~m~p$.
If the license starts with  $\Gfor~m~p$,
then the body of the license is valid for $m$ time periods of length
$p$, but can be canceled at the end of any period.

As an example, 
consider 
the license
\[\Gfor~3~100~\Gpay~10.00~\Gflatrate~\Gfor~W~\Gon~D\]
where $W$ is a set of works and $D$ is a set of devices.  This license
allows the client to render 
any work in $W$ on a device in $D$ by paying a flat rate of $10.00$ at 
the end of every $100$ time units, for $3$ such time periods. 

We can incorporate this license language in our logic by replacing our
syntax for licenses ($\ell$) with expressions in the above
language. To define the function $\fnL{-}$, which interprets licenses as
sets of traces in the semantics of our logic, we adapt the semantics
of \cite{Gunter01}.  
(The main difference is that we have a fixed time granularity, whereas the 
original semantics uses real numbers as time stamps for events.) 

To build up the function $\fnL{-}$, we first assign 
sets of
traces to the simplest licenses, those that are valid for a
single period. 
The set of traces that allow for a payment of $x$ to view works from $W$ on 
devices from $D$, for a period of $p$ time units depends on the payment schedule.
The traces for an up front schedule is defined as:
\ifCorr
\begin{align*}
\mathit{UpFront}(x,p,W,D)  = 
 \{ \mathsf{pay}[x] a_1\cdots a_{p-1} ~|~ & 
\mbox{$a_i$ is either $\bot$ or $\mathsf{render}[w,d]$}\\
& \mbox{for some $w\in W$ and $d\in D$}\}.
\end{align*}
\else
\[\begin{array}{l}
\mathit{UpFront}(x,p,W,D)  = \\
\qquad  \{ \mathsf{pay}[x] a_1\cdots a_{p-1} ~|~ \\
\qquad\quad\mbox{$a_i$ is either $\bot$ or $\mathsf{render}[w,d]$}\\
\qquad\quad\mbox{for some $w\in W$ and $d\in D$}\}.
  \end{array}\]
\fi
\commentout{
Similarly, the set of traces that allow for a flat rate payment of $x$
to view works from $W$ on devices from $D$, for a period of $p$ time
units is defined as: 
}
The traces for a flat rate schedule is defined as:
\ifCorr
\begin{align*}
\mathit{FlatRate}(x,p,W,D)  = 
 \{ a_0\cdots a_{p-2}\mathsf{pay}[x] ~|~ & 
\mbox{$a_i$ is either $\bot$ or $\mathsf{render}[w,d]$}\\
& \mbox{for some $w\in W$ and $d\in D$}\}.
\end{align*}
\else
\[\begin{array}{l}
\mathit{FlatRate}(x,p,W,D)  = \\
\qquad \{ a_0\cdots a_{p-2}\mathsf{pay}[x] ~|~ \\
\qquad\quad\mbox{$a_i$ is either $\bot$ or $\mathsf{render}[w,d]$}\\
\qquad\quad\mbox{for some $w\in W$ and $d\in D$}\}.
  \end{array}\]
\fi
\commentout{
Finally, the set of traces that allow for a payment of $x$
per use of a work $w$ from $W$ on devices from $D$ during a period of
$p$ time units is defined as:
}
The
set of traces for a per use schedule is defined as:
\ifCorr
\begin{align*}
\mathit{PerUse}(x,p,W,D) = 
\{ a_0\cdots a_{p-2}\mathsf{pay}[nx] ~|~ &
\mbox{$a_i$ is either $\bot$ or $\mathsf{render}[w,d]$}\\
& \mbox{for some $w\in W$ and $d\in D$,}\\ 
& \mbox{and $n=|\{a_i~|~a_i\not=\bot\}|$}\}.
\end{align*}
\else
\[\begin{array}{l}
\mathit{PerUse}(x,p,W,D) = \\
\qquad\{ a_0\cdots a_{p-2}\mathsf{pay}[nx] ~|~ \\
\qquad\quad\mbox{$a_i$ is either $\bot$ or $\mathsf{render}[w,d]$}\\
\qquad\quad\mbox{for some $w\in W$ and $d\in D$,}\\ 
\qquad\quad\mbox{and $n=|\{a_i~|~a_i\not=\bot\}|$}\}.
  \end{array}\]
\fi
Given two sets of traces $S_1$ and $S_2$, we define $S_1\cdot S_2$ as
the set $\{s_1 \cdot s_2 ~|~ s_1\in S_1, s_2\in S_2\}$. In other
words, $S_1\cdot S_2$ is the set of all concatenation of traces from
$S_1$ and $S_2$. We write $S^n$ for $\underbrace{S\cdot S\cdot\ldots\cdot
S}_{n}$. 

Using the above definitions, we define the function $\fnL{-}$ as: 
\begin{eqnarray*}
\fnL{\Gfor~p~z} & = & \fnM{z}(p)\\
\fnL{\Gfor~m~p~z} & = & (\fnM{z}(p))^m\\
\fnL{\Gfor~\Gupto~m~p~z} & = & \bigcup_{n=0}^{m}(\fnM{z}(p))^n,
\end{eqnarray*}
where $\fnM{-}$ generates the traces for a single time period: 
\[\begin{array}{l}
\fnM{\Gpay~x~\Gupfront~\Gfor~W~\Gon~D}(p) = 
  \mathit{UpFront}(x,p,W,D)\\
\fnM{\Gpay~x~\Gflatrate~\Gfor~W~\Gon~D}(p) =
  \mathit{FlatRate}(x,p,W,D)\\
\fnM{\Gpay~x~\Gperuse~\Gfor~W~\Gon~D}(p) =
  \mathit{PerUse}(x,p,W,D).
  \end{array}\]
As expected, the semantics of the logic defined in
Section~\ref{s:logic} carries over verbatim with the above
changes. 

The \emph{DigitalRights} language given above is not more expressive
than the regular one that we introduced in Section~\ref{s:logic}.  It
is easy to see that for any license $e$ in \emph{DigitalRights}, the
set of traces $\fnL{e}$ can be expressed by a regular
language. Because the sets $\mathit{UpFront}(x,p,W,D)$,
$\mathit{FlatRate}(x,p,W,D)$, and $\mathit{PerUse}(x,p,W,D)$ are
finite for any $p$, $x$, $W$ and $D$, it is trivial to express them
using a regular language. The concatenation operation $S_1\cdot S_2$
preserves regularity, as does union, therefore it is possible to
express any license expressed in \emph{DigitalRights} as a regular
one.  There are, however, advantages to using the \emph{DigitalRights} 
language. The translation of a \emph{DigitalRights} license yields a
large regular expression that may be significantly less efficient to
verify than the original license. Another benefit is that the
\emph{DigitialRights} language is easier to understand.

It should be noted that every license language is not necessarily
subsumed by the language of regular expressions. To see this, consider
a license in some license language that can be canceled whenever the
number of renderings equals the number of payments. The set of traces
corresponding to such  a license is not regular, by a well-known
result from formal language theory (see for instance
\cite{Hopcroft69}). 
Therefore, any language that can be used to state this license is not 
equivalent to any sublanguage of the regular expressions.

\section{Related work}\label{s:related}

The inspiration for our work comes from the field of program
verification, where one finds logics such as Hoare Logic
\cite{Hoare69} 
and 
Dynamic Logic \cite{Harel00} to reason about
properties of programs.  Our logic is similar to those, in the sense
that our formulas contain explicit licenses, in much the same way that 
theirs contain explicit programs. Logics of this type are often
referred to as \emph{exogenous}. In contrast, \emph{endogenous} logics 
do not explicitly mention programs; to analyze a program with such a
logic, one builds a model for that specific program, and uses the
logic to analyze 
the
model. One advantage of using an exogenous logic
is that it allows 
the behavior of two programs to be compared
within the logic. In our  case, it allows us to compare the effect of
different licenses within the logic. 
An endogenous logic, however, permits
more efficient verification
procedures. 
To get this benefit,
our verification procedures in
Section~\ref{s:verification} essentially convert formulas from our
logic into formulas of an endogenous logic, viz. temporal logic. 

Although our logic is an exogenous logic inspired by Dynamic Logic, its
models are quite different. In Dynamic Logic, programs guide the state 
transitions in the model. Licenses, on the other hand, do not affect
states. Instead, they are used to specify permissions and
obligations. The models of our logic are primarily influenced by the
work of Halpern and van der Meyden \citeyear{Halpern01a} on
formalizing SPKI \cite{spki2}. 
SPKI is used to account for
access rights based on certificates received. Similarly, we base the
right to do actions on the licenses received. In fact, we could
imagine licenses being implemented with SPKI certificates. 

Permissions and obligations are key concepts in our approach. These
notions are typically studied in the philosophical literature under
the heading of \emph{deontic logic} \cite{Meyer93}. Early accounts of
deontic logic failed to differentiate between actions and assertions,
leading to many paradoxical and counterintuitive propositions (see for
instance \cite{Follesdal81}). The idea of separating actions from
assertions has lead to a recasting of deontic logic as a variant of
Dynamic Logic \cite{Meyer88,Meyden90}.  
Models for deontic dynamic logics
specify explicitly 
either 
which states represent the violation of an
obligation or a permission
or which transitions are permitted or forbidden. In \cite{Meyer88}, 
a special formula $V$ is introduced in the
logic, and any state that satisfies $V$ is deemed a
violation. Intuitively, an action $a$ is permitted in a state if it is
possible to reach a state via $a$ where $V$ does not hold. Conversely,
an action is obligatory if performing any other action leads to a
state where $V$ holds. 
In \cite{Meyden90}, it is the transitions between states that are
deemed permitted or forbidden. 
$\llic$ is different from these approaches, because
we derive our permissions and obligations
from  the licenses issued in the run. This indirection means that we
do not have to explicitly model the permissions and obligations.
In addition, we can easily change the model to account for different 
licenses.  

Finally, deontic logic has been used to reason about contracts. This
is intriguing, because a license can be viewed as a  restricted form
of contract. Research in this direction includes work by Lee 
\citeyear{Lee88a}, which focuses on developing a logical language
based on predicate logic with temporal operators. Deontic operators
are handled using a specific predicate to represent a violation (in
this context, defaulting on a contract). Unfortunately, the logic is
not meant to reason about contracts written in some language. Instead, 
the models for the logic \emph{represent} the contracts to be
analyzed. In other words, for each contract that he wants to study,
Lee builds a specific model encoding violations at the appropriate
states.

\section{Conclusion}

In this paper we have introduced a framework for precisely stating and
rigorously proving properties of licenses.  We also have illustrated
how our logic can be modified to reason about licenses that are
written in any language with a trace-based semantics.  This
flexibility provides us with a common ground in which to compare
different rights languages with trace-based semantics.  We intend to
report on these comparisons in the future.  While useful in its own
right, the logic is a simple foundation on which more expressive
rights management logics can be built.  For example, the logic can be
modified in a straightforward manner to support multiple clients and
multiple providers. Multiple providers is an especially interesting
case, because it allows us to study the management of licensing
rights, the rights required for one provider to legitimately offer
another provider's work to a client.  We plan to examine various
extension in the near future. 
There remain interesting questions about the foundation of $\llic$,
such as axiomatizations for the logic.
Finally, as mentioned previously, our
operators $P$ and $O$ have a distinctly deontic flavor. It would be
interesting to establish a correspondence between our approach and
existing deontic frameworks, in particular deontic logics of actions
\cite{Khosla87,Meyer88,Meyden90}.

\section*{Acknowledgments}

We would like to thank Joe Halpern, Carl Lagoze and Sandy Payette for their
helpful comments. 
Joshua Guttman and the CSFW anonymous referees made
suggestions that greatly improved the presentation. 
Support for this
work came from the ONR under grants N00014-00-1-0341 and
N00014-01-1-0511, 
from the DoD Multidisciplinary University Research
Initiative (MURI) program administered by the ONR under
grant N00014-01-1-0795, 
and from the  NSF under grant IIS-9905955 (Project Prism).  

\appendix

\section{Proofs}

\opro{p:validp}
For all action expressions $(a,n)$, the formula $P(a,n)\vee
P(\overline{a},n)$  is valid.
\eopro
\prf The validity of this formula is a consequence of the fact that
$P_r(t)$ contains at least one action corresponding to every license
name $n$. Given a run $r$ and a time $t$, and consider the action
expression $(a,n)$. We know there must exist an action-name pair
$(b,n)$ in $P_r(t)$. Two cases arise. If $a=b$, then $(a,n)$ is in
both $\fnA{(a,n)}$ and $P_r(t)$, and thus $r,t\sat P(a,n)$. If
$a\not=b$, then $(b,a)$ is in both $\fnA{(\overline{a},n)}$ and
$P_r(t)$, and thus $r,t\sat P(\overline{a},n)$. Therefore, we have
$r,t\sat P(a,n)\vee P(\overline{a},n)$. Since the above holds for all
$r$ and $t$, $\sat P(a,n)\vee P(\overline{a},n)$. \eprf  

\opro{p:runencoding}
If $r$ is a finite run and $N_\phi\subseteq N_r$, 
then $r,t \sat \phi$ iff $\sat\psi_r\rimp \Circ^t\phi$.
\eopro

To simplify the proof, we introduce the following notation. Given 
runs $r,r'$, times $t,t'$, and a subset $N$ of $\Names$, define
$(r,t)\leq_N(r',t')$ if for all $i\geq 0$,
$\mathit{lic}(r,t+i)\subseteq\mathit{lic}(r',t'+i)$ and
$\left(\mathit{act}(r,t+i)\cap(\Act\times N)\right) =
\left(\mathit{act}(r',t'+i)\cap(\Act\times N)\right)$. Intuitively,
$(r,t)\leq_N(r',t')$ if every license issued by $r$ (starting at time
$t$) is also issued in $r'$ (starting at time $t')$, and moreover the
two runs agree on the actions corresponding to license names in
$N$. The following lemmas capture the relevant properties of the
$\leq_N$ relation. Recall that $N_\phi$ is the set of license names
appearing in formula $\phi$. 
\lem\label{l:runencoding}
For any $\phi$ such that $N_\phi\subseteq N_r$, if
$(r,0)\leq_{N_r}(r',t')$, then $r,i\sat \phi$ iff $r',t'+i\sat\phi$ for
all $i\geq 0$. 
\elem
\prf By induction on the structure of $\phi$. We prove the nontrivial
cases here. Consider $\phi=n:\ell$. If $r,i\sat n:\ell$, then
$(n,\ell)\in\mathit{lic}(r,i)\subseteq\mathit{lic}(r',t'+i)$, and hence
$r',t'+i\sat n:\ell$. Conversely, if $r',t'+i\sat n:\ell$, then since
$N_\phi\subseteq N_r$, license name $n$ must appear in $r$, and by
definition of $(r,0)\leq_{N_r}(r',t')$ and the fact that license names 
can be associated with only one license in a run, it must be the case
that $(n,\ell)\in\mathit{lic}(r,i)$. Hence, $r,i\sat n:\ell$. The
cases for $(a,n)$ and $(\overline{a},n)$ follow from $r$ and $r'$
agreeing on the actions for license names $n\in N_\phi\subseteq
N_r$. For $P(a,n)$ and $P(\overline{a},n)$, because $r$ and $r'$ agree 
on the licenses issued with name $n\in N_\phi\subseteq N_r$, and
because $r$ and $r'$ agree on the actions pertaining to license name
$n$, $P_r$ and $P_{r'}$ agree on the permissions with respect to
license name $n$, from which the result follows. The remaining cases
are a straightforward application of the inductive hypothesis. \eprf

\lem\label{l:runencoding2}
$r',t'\sat\psi_r$ iff $(r,0)\leq_{N_r}(r',t')$.
\elem
\prf We know by definition that $r',t'\sat\psi_r$ if and only if
$r',t'\sat\psi_0$, $r',t'+1\sat\psi_1$, $\ldots$,
$r',t'+t_f\sat\psi_{t_f}$, and $r,t'+t\sat\psi_e$ for all
$t>t_f$. Given the definition of $\psi_0,\ldots,\psi_{t_f}$ and
$\psi_e$, this is equivalent to
$\mathit{lic}(r,0)\subseteq\mathit{lic}(r',t')$, $\ldots$,
$\mathit{lic}(r,t_f)\subseteq\mathit{lic}(r',t'+t_f)$,
$\mathit{lic}(r,t)=\emptyset\subseteq\mathit{lic}(r',t'+t)$ for
$t>t_f$, and moreover $r(i)$ and $r'(t'+i)$ agree on the actions
pertaining to license names $n\in N_r$ for all $i\geq 0$. This just
says that $(r,0)\leq_{N_r}(r',t')$. \eprf

\prf (Proposition~\ref{p:runencoding}) Note that $r,t\sat\phi$ iff
$r,0\sat\Circ^t\phi$. Thus, it is sufficient to show that
$r,0\sat\phi$ iff $\sat\psi_r\rimp\phi$. 

First, assume that
$(r,0)\sat\phi$. Let $r',t'$ be an arbitrary run and time. If
$r',t'\sat\psi_r$, then by Lemma~\ref{l:runencoding2},
$(r,0)\leq_{N_r}(r',t')$. Since $N_\phi\subseteq N_r$,
Lemma~\ref{l:runencoding} implies that $r',t'\sat\phi$. This
establishes that $r',t'\sat\psi_r\rimp\phi$. Since $r',t'$ was
arbitrary, $\sat\psi_r\rimp\phi$ holds. 

For the converse direction, assume that $\sat\psi_r\rimp\phi$. In
particular, $r,0\sat\psi_r\rimp\phi$. Since $(r,0)\leq_{N_r}(r,0)$,
Lemma~\ref{l:runencoding2} implies that $r,0\sat\psi_r$, and hence
$r,0\sat\phi$. \eprf

\opro{p:licencoding}
For any license $\ell$, the formulas $n:\ell\rimp\phi_{n,\ell}^i$ are
valid, for $i=0,1,2,\ldots$. 
\eopro
\prf The proof relies on a suitable application of standard properties 
of regular expressions, and much formal symbolic manipulation. We
sketch the argument here. First, extend the definition of $S$ to 
handle more than a single action. 
Let $S^k(\ell)$ (for $k\geq 1$) be the
function that returns the set of all prefixes of 
length $k$ of action sequences associated with $\ell$. Formally,
$S^1(\ell)=S(\ell)$, and $S^{k+1}=\{a\sigma ~:~ a\in S(\ell), \sigma\in 
S^k(D_a(\ell))\}$. 

Given this definition, we can verify that the formula
$\phi_{n,\ell}^{i+1}$ is equivalent to
$\phi_{n,\ell}^i\wedge\phi_{n,\ell}^{i\mapsto i+1}$, where
$\phi_{n,\ell}^{i\mapsto i+1}$ is the formula
\[\bigwedge_{a_0\cdots a_{i+1}\in \atop S^{i+2}(\ell)}\left(\begin{array}{l}
        \left((a_0,n)\wedge\Circ(a_1,n)\wedge\cdots\right.\\
              \left.\wedge\Circ^i(a_i,n)\right)\rimp\Circ^{i+1}P(a_{i+1},n) \end{array}
\right).\]

Let $r,t$ be an arbitrary run and time. We show by induction that
$r,t\sat n:\ell\rimp\phi_{n,\ell}^i$ for all $i\geq 0$. Assume
$r,t\sat n:\ell$, that is, $(n,\ell)\in\mathit{lic}(r,t)$. The base
case of the induction is verified by noticing that $\phi_{n,\ell}^0 =
\bigwedge_{a\in S(\ell)} P(a,n)$, and by the definition of $P_r(t)$,
for all $a\in S(\ell)$, $(a,n)\in P_r(t)$, so that $r,t\sat
P(a,n)$. The induction step follows by a similar reasoning. Assume
$r,t\sat\phi_{n,\ell}^i$. Given the above equivalence, it is
sufficient to show that $r,t\sat\phi_{n,\ell}^{i\mapsto i+1}$ to
establish the result. For any $a_0\cdots a_{i+1}\in S^{i+2}(\ell)$,
if $r,t\sat(a_0,n)\wedge\Circ(a_1,n)\wedge\cdots\wedge\Circ^i(a_i,n)$, 
then $r,t\sat(a_0,n)$, $r,t+1\sat(a_1,n)$, $\ldots$,
$r,t+i\sat(a_i,n)$. Since $a_0\cdots a_i a_{i+1}\in S^{i+2}(\ell)$, it 
is viable for $\ell$, and hence $(a_{i+1},n)\in P_r(t+i+1)$, that is,
$r,t+i+1\sat P(a_{i+1},n)$, or $r,t\sat\Circ^{i+1}P(a_{i+1},n)$, as
required. Since this is true for all sequences in $S^{i+2}(\ell)$,
we have $r,t\sat\phi_{n,\ell}^{i\mapsto i+1}$, establishing our result.
\eprf

\opro{p:ltlmodelc}
$r,t\sat \phi$ iff $M_r,s_t \sat_L \phi^T$.
\eopro
\prf  We prove by induction on the structure of $\phi$ that for all
$t$, $r,t\sat\phi$ iff $M_r,s_t\sat_L \phi^T$. We give a few
representative cases here, the remaining cases being similar.

Consider $\phi=n:\ell$. For any $t$, we have $r,t\sat n:\ell$ iff
$(n,\ell)\in\mathit{lic}(r,t)$ iff $\mathsf{issued}(n,\ell)\in L(s_t)$ (by
construction of $L(s_t)$) iff $M_r,s_t\sat_L\mathsf{issued}(n,\ell)$. 

Consider $\phi=P(\overline{a},n)$. For any $t$, we have $r,t\sat
P(\overline{a},n)$ iff $(b,n)\in P_r(t)$ for some $b\not=a$ iff
$\mathsf{obligated}(a,n)$ is \emph{not} in $L(s_t)$ (since $(a,n)$
cannot be the unique action in $P_r(t)$) iff
$M_r,s_t\sat_L\neg\mathsf{obligated}(a,n)$. 

Consider $\phi=\Circ\phi'$. For any $t$, we have $r,t\sat\Circ\phi'$
iff $r,t+1\sat\phi'$ iff $M_r,s_{t+1}\sat_L(\phi')^T$ (by hypothesis) iff $M_r,s_t\sat_L \mathbf{X}(\phi')^T$, and $\mathbf{X}(\phi')^T = \phi^T$. 
\eprf

\opro{p:licmodelc}
If $M,s\sat_L\phi^T\wedge\phi^I$, then there exists a run $r$ such that 
$r,0\sat\phi$. 
\eopro
\prf Without loss of generality, $M=(S,L)$ with $S=\{s_0,s_1,\ldots\}$, 
and $s=s_0$. (If not, $s=s_t$ for some $t$, and take $M'=(S',L)$ where
$S'=\{s_t,s_{t+1},\ldots\}$, and we can check that
$M',s_0\sat_L\phi^T\wedge\phi^I$.) Construct the run $r_M$ as 
follows: for all $t\geq 0$, $r_M(t)=(L_M(t),A_M(t))$, where $L_M(t)=\{
(n,\ell) ~:~ \mathsf{issued}(n,\ell)\in L(s_t)\}$, and $A_M(t)(n)=a$
if $\mathsf{done}(a,n)\in L(s_t)$, and $A_M(t)(n)=\bot$
otherwise. This is a well-defined run, because $M_r,s_0$ satisfies 
$\mathsf{Done}_{A_\phi,N_\phi}$ and
$\mathsf{Issued}_{L_\phi}$. We next check that for all $t\geq 0$,
$P_{r_M}(t)=\{ (a,n) ~:~ \mathsf{permitted}(a,n)\in L_M(s_t)\}$. The
details are routine, if tedious. Essentially, every path through the
automaton encoded in $\mathsf{NFA_{n,\ell,A_\phi}}$ corresponds to a
viable trace of the license $\ell$ from the point where the license is 
issued. A straightforward proof by induction establishes that
$r_M,0\sat\phi$. 
\eprf

\othm{t:satisfiability}
The satisfiability problem for $\llic$ is PSPACE-complete.
\eothm
\prf For the lower bound, we show that we can reduce the
satisfiability problem for LTL to the satisfiability problem for
$\llic$. Let $F$ be a formula of LTL, over primitive propositions
$\Phi_f=\{p_1,\ldots,p_n\}$. We first 
rewrite $F$ into a formula $\phi_F$ of $\llic$, by picking an
arbitrary non-$\bot$ action in $\Act$ (call it $\star$) and a name $n_p$ for every $p\in\Phi_f$, and replacing every
primitive proposition $p$ in $F$ by the action expression $(\star,n_p)$,
and replacing $\mathbf{G}$, $\mathbf{X}$, and $\mathbf{U}$ by $\Box$,
$\Circ$, and $\cU$ respectively. Assume $F$ is satisfiable in a
linear structure $M=(S,L)$ at state $s_i$, where $S=(s_0,s_1,\ldots)$. Let $r_M$ be the run
defined by $r_M(t) = (\emptyset,A(t))$, where $A(t)$ maps name $n_p$ to
action $\star$ if $p\in L(s_t)$, and to $\bot$ otherwise, and maps all
other names to $\bot$. It is easy to check that $\phi_F$ is
satisfiable in $r_M$ at time $i$. Similarly, if $\phi_F$ is
satisfiable in a run $r$ at time $t$, we can convert $r$ into a linear 
structure $M_r=(S,L)$, where $p\in L(s_t)$ iff $(\star,n_p)\in\mathit{act}(r,t)$,
and it is easy to check that $F$ is satisfiable in $M_r$ at state
$s_t$. Since the satisfiability problem for LTL is PSPACE-complete,
the above reduction means that the satisfiability problem for $\llic$
is PSPACE-hard.

For the upper bound, we show that we can reduce the satisfiability
problem for $\llic$ to the satisfiability problem for LTL in
polynomial time. In
particular, we show that $\phi$ is satisfiable in $\llic$ iff
$\phi^T\wedge\phi^I$ is satisfiable in LTL. Let $\phi$ 
be a formula satisfied in run $r$ at time $t$. By Proposition~\ref{p:ltlmodelc},
$M_r,s_t\sat_L\phi^T$. 
By
construction,  it is clear that $M_r,s_t\sat_L\phi^I$ (only one action
per license per time, no two licenses with the same name ever issued,
and so on).  Hence,
$M_r,s_t\sat_L\phi^T\wedge\phi^I$. Conversely, assume that
$\phi^T\wedge\phi^I$ is satisfiable in a linear structure
$M$. By Proposition~\ref{p:licmodelc}, there exists a run $r$
such that $r,0\sat\phi$, i.e., $\phi$ is satisfiable in
$\llic$. Finally, one can check that the 
size of the formula $\phi^T\wedge\phi^I$ is polynomial in
the size of $\phi$. 
\eprf

\opro{p:minimalint}
There exists a polynomial time algorithm for computing the 
interpretation $P_r$ corresponding to a finite run $r$. 
\eopro

\prf It is clearly sufficient to
define $P_r$ for non-$\bot$ actions only, by taking $\bot$ to be the
default value of $P_r$.  Let $L_r$ be the set of named licenses issued
in run $r$. We define, for every named license $(n,\ell)\in L_r$, a
function $P_{r,n}$ that gives for every time $t$ the set of actions
permitted by the named license $(n,\ell)$ at time $t$. Clearly, we can 
then take $P_r(t)=\bigcup_{(n,\ell)\in L_r}P_{r,n}(t)$. 

Consider a named license $(n,\ell)\in L_r$, and assume $(n,\ell)$ is
issued at time $t_0$ in $r$. Let $A=(Q,I,\Delta,F)$ be the
$\epsilon$-free NFA corresponding to the regular expression $\ell$,
where $Q$ is the set of states, $I$ is the set of initial states,
$\Delta$ is the transition relation, and $F$ is the set of final
states. We can construct $A$ in time polynomial in the size of $\ell$,
using \cite{Hromkovic97}, where $|Q|$ is linear in the size of $\ell$
and $|\Delta|$ is less than quadratic.

We can now define the function $P_{r,n}$. For $t<t_0$, we can take
$P_{r,n}(t)=\{\bot\}$. For $t\geq t_0$, we need to take the license
into consideration. First, define the sequence of sets
$S_0,S_1,\ldots,S_{m-t_0}$ where $m$ is the length of run $r$. These
sets represents the sets of states of the NFA obtained by following
the actions related to license name $n$ prescribed by the
run. Formally, define $S_i$ inductively as:
\begin{eqnarray*}
S_0 & = & I\\
S_{i+1} & = & \{s' ~:~ \mbox{$(s,a,s')\in\Delta$ for some} \\
 & & \quad\qquad\mbox{$s\in S_i$ and $(a,n)\in\mathit{act}(r,t_0+i)$} \}.
\end{eqnarray*}
With these sets, we define $P_{r,n}(t_0+i)=\bigcup_{s\in S_i}\{a ~:~ \exists
s'.(s,a,s')\in\Delta\}$, that is, the set of actions that can be
performed according to license $\ell$ starting from any of the states
in $S_i$. One can check that the sets $S_i$ can be constructed in
polynomial time, and therefore that $P_{r,n}$, and hence $P_r$, can be
constructed in polynomial time. 
\eprf

\othm{t:modelchecking}
There exists an algorithm for deciding if a formula $\phi$ is true in a 
finite run $r$ at time $t$.  Furthermore, the algorithm runs in polynomial
time with respect to the size of the model $r$ and in exponential time with
respect to the size of the formula $\phi$.
\eothm

\prf  Given a run $r$, we can compute $P_r$ in polynomial time by
Proposition~\ref{p:minimalint}, and construct the model $M_r$ in time
polynomial in the size of $r$. We can translate $\phi$ into $\phi^T$
in time polynomial in the size of the formula. We use
Proposition~\ref{p:ltlmodelc} to reduce the problem to  the
model-checking problem for LTL, which can be solved in time polynomial
in the size of the $M_r$ and exponential in the size of $\phi$ (see,
for instance, \cite{Vardi97}). 
\eprf

\opro{p:ltlvalc}
$r\sat\phi$ iff $M_r,s_0\sat_L \mathbf{G}(\phi^T)$.
\eopro
\prf By definition, $r\sat\phi$ iff for all times $t$,
$r,t\sat\phi$. By Proposition~\ref{p:ltlmodelc}, this holds iff for all 
states $s_t$ of $M_r$, $M_r,s_t\sat \phi^T$, which just means that
$M_r,s_0\sat \mathbf{G}\phi^T$. \eprf

\ifCorr
 \bibliographystyle{chicagor}
\else
\bibliographystyle{latex8}
\fi
\bibliography{riccardo,joe}

\end{document}